\documentstyle[12pt]{article}
\def\be{\begin{equation}}
\def\ee{\end{equation}}
\def\bc{\begin{center}}
\def\ec{\end{center}}
\def\bea{\begin{eqnarray}}
\def\eea{\end{eqnarray}}
\def\atu{{\alpha_t^U}}
\def\cm{{\cal M}}
\def\dvc{{\Delta V_{cosm}}}
\def\ov{\overline}
\def\gev{{\rm \; GeV}}

\def\mpl{M_{\rm P}}
\def\msu{{M_{\rm SUSY}}}

\catcode`@=11
\def\marginnote#1{}
\newcount\hour
\newcount\minute
\newtoks\amorpm
\hour=\time\divide\hour by60
\minute=\time{\multiply\hour by60 \global\advance\minute by-\hour}
\edef\standardtime{{\ifnum\hour<12 \global\amorpm={am}%
        \else\global\amorpm={pm}\advance\hour by-12 \fi
        \ifnum\hour=0 \hour=12 \fi
        \number\hour:\ifnum\minute<10 0\fi\number\minute\the\amorpm}}
\edef\militarytime{\number\hour:\ifnum\minute<10 0\fi\number\minute}
\def\draftlabel#1{{\@bsphack\if@filesw {\let\thepage\relax
   \xdef\@gtempa{\write\@auxout{\string
      \newlabel{#1}{{\@currentlabel}{\thepage}}}}}\@gtempa
   \if@nobreak \ifvmode\nobreak\fi\fi\fi\@esphack}
        \gdef\@eqnlabel{#1}}
\def\@eqnlabel{}
\def\@vacuum{}
\def\draftmarginnote#1{\marginpar{\raggedright\scriptsize\tt#1}}
\def\draft{\oddsidemargin 0.0truein
        \def\@oddfoot{\sl preliminary draft \hfil
        \rm\thepage\hfil\sl\today\quad\militarytime}
        \let\@evenfoot\@oddfoot \overfullrule 3pt
        \let\label=\draftlabel
        \let\marginnote=\draftmarginnote
   \def\@eqnnum{(\theequation)\rlap{\kern\marginparsep\tt\@eqnlabel}%
\global\let\@eqnlabel\@vacuum}  }
\catcode`@=12
\begin{document}
\begin{titlepage}
\vspace*{-1cm}
\phantom{bla}
\hfill{CERN-TH.7185/94}
\\
\phantom{bla}
\hfill{LPTENS-94/08}
\\
\phantom{bla}
\hfill{hep-ph/9406256}
\vskip 1.5cm
\begin{center}
{\Large\bf Towards a dynamical determination of parameters}
\end{center}
\begin{center}
{\Large\bf in the Minimal Supersymmetric
Standard Model}\footnote{Work
supported in part by the US NSF under
grant No.~PHY89--04035, and by the European
Union under contract No.~CHRX-CT92-0004.}

\end{center}
\vskip 1.0cm
\begin{center}
{\large Costas Kounnas}\footnote{On leave from the Laboratoire de
Physique Th\'eorique, ENS, Paris, France.}, {\large Fabio
Zwirner}\footnote{On leave from INFN, Sezione di Padova, Padua,
Italy.}
\\
\vskip .1cm
Theory Division, CERN, \\
CH-1211 Geneva 23, Switzerland
\\
\vskip .3cm
and
\\
\vskip .3cm
{\large Ilarion Pavel}
\\
\vskip .1cm
Laboratoire de Physique Th\'eorique, ENS, \\
F-75231 Paris Cedex 05, France. \\
\end{center}
\vskip 0.5cm
\begin{abstract}
\noindent
In the Minimal Supersymmetric Standard Model (MSSM), the scale $\msu$
of soft supersymmetry breaking is usually {\em assumed} to be of the
order of the electroweak scale. We reconsider here the possibility of
treating $\msu$ as a dynamical variable. Its expectation value should
be determined by minimizing the vacuum energy, after including MSSM
quantum corrections. We point out the crucial role of the
cosmological
term for a dynamical generation of the desired hierarchies $m_Z, \msu
\ll \mpl$. Inspired by four-dimensional superstring models, we also
consider the Yukawa couplings as dynamical variables. We find that
the top Yukawa coupling is attracted close to its effective infrared
fixed point, corresponding to a top-quark mass in the experimentally
allowed range.  As an illustrative example, we present the results of
explicit calculations for a special case of the MSSM.
\end{abstract}
\vfill{
CERN-TH.7185/94
\newline
\noindent
May 1994}
\end{titlepage}
\setcounter{footnote}{0}
\vskip2truecm
\vspace{1cm}
{\bf 1.}
The most plausible solution of the naturalness or hierarchy problem
of the Standard Model is low-energy supersymmetry, whose simplest
realization is [\ref{mssm}] the Minimal Supersymmetric extension of
the Standard Model (MSSM). Such a choice is sufficient to guarantee
the quantum stability of the gauge hierarchy $m_Z \ll M$, where $M$
is the appropriate ultraviolet cut-off scale for the MSSM,
somewhere\footnote{In models where supersymmetry is spontaneously
broken by non-perturbative phenomena such as gaugino condensation,
the appropriate cut-off scale might be different, but it is in any
case much larger than the electroweak scale.} near the
grand-unification
scale $M_U \simeq 2 \times 10^{16} \gev$ and the Planck scale $\mpl
\equiv G_N^{-1/2} / \sqrt{8 \pi} \simeq 2.4 \times 10^{18} \gev$.
In addition to the stabilization of the hierarchy, another very
attractive feature of the MSSM is the possibility of describing
the spontaneous breaking of the electroweak gauge symmetry as an
effect of radiative corrections [\ref{radbr}].

Besides these virtues, the MSSM has a very unsatisfactory feature:
the numerical values of its explicit mass parameters must
be arbitrarily chosen `by hand' (the fact that also the gauge and
Yukawa couplings must be chosen `by hand' is as unsatisfactory as in
the Standard Model). This means a certain lack of predictivity, and
in particular does not provide any dynamical explanation for the
origin of the hierarchy $\msu \ll \mpl$, which is just assumed to be
there. To go further, one must have a model for spontaneous
supersymmetry breaking in the fundamental theory underlying the MSSM.
The only possible candidate for such a theory is $N=1$ supergravity
coupled to gauge and matter fields [\ref{cfgvp}], where (in contrast
with the case of global supersymmetry) the spontaneous breaking of
local supersymmetry is not incompatible with vanishing vacuum energy.
In realistic supergravity models,
supersymmetry breaking typically occurs in a `hidden sector', whose
interactions with the MSSM states are of gravitational strength. For
spontaneous breaking on a flat background, the order parameter is
the gravitino mass, $m_{3/2}$, and all the explicit mass parameters
of the MSSM are calculable [\ref{bfs}] (but model-dependent)
functions
of $m_{3/2}$.

The general structure of $N=1$ supergravity cannot provide
significant information on the MSSM parameters, but allows us to
address some very important questions, which, irrespectively of the
specific mechanism
that breaks supersymmetry, can lead to important constraints, and
hopefully guide us towards the identification of a more fundamental
theory. Among the possible
questions, we would like to concentrate here on the following ones:
1) how to avoid a large cosmological constant, already at the
classical level; 2) how to generate the hierarchy $m_{3/2} \ll \mpl$
without explicitly introducing small mass parameters; 3) how to avoid
that quadratically divergent loop corrections to the effective
potential generate an unacceptably large cosmological constant
${\cal O}(m_{3/2}^2 \mpl^2)$ and possibly destabilize the hierarchy
$m_{3/2} \ll \mpl$.

The first two questions were the main motivations of the so-called
{\em no-scale} supergravity models [\ref{cfkn}, \ref{ekn}]. Their
main feature is that the vacuum energy is equal to zero [or at most
${\cal O} (m_{3/2}^4)$] at the classical level, even after
supersymmetry breaking, thanks to the existence of at least one flat
[or almost flat] direction in the scalar potential [\ref{cfkn}].
The gravitino mass $m_{3/2}$, which sets the scale $M_{\rm SUSY}$ of
the soft masses, is not fixed at the classical minimum, but is a
function
of the scalar field(s) parametrizing the flat [or almost flat]
direction(s).
If quantum-gravity effects at the Planck scale do not change
drastically
the approximate flatness of the scalar potential, then $m_{3/2}$ is
determined by the quantum corrections associated with the light MSSM
fields [\ref{ekn}]. Within this class of models, there exists a
possibility of explaining the hierarchy by minimizing the effective
potential not only with respect to the Higgs fields, but also with
respect to $M_{\rm SUSY}$: as we shall see in detail later, there are
situations in which $m_Z$ and $M_{\rm SUSY}$ are both dynamically
determined to be exponentially suppressed with
respect to $\mpl$.

In a generic $N=1$ supergravity, the viability of the above scenario
(and of other scenarios for the generation of the hierarchy $m_{3/2}
\ll \mpl$) is plagued by the possible existence of quadratically
divergent one-loop contributions to the vacuum energy, proportional
to ${\rm Str \,} \cm^2 \equiv \sum_i (-1)^{2 J_i} (2 J_i + 1) m_i^2$,
where $m_i$ and $J_i$ are the field-dependent mass-eigenvalue and the
spin of the $i$-th particle, respectively.
In the fundamental quantum theory of gravity, these corrections would
correspond to finite contributions to the effective potential, ${\cal
O} (m_{3/2}^2 \mpl^2)$. Since $m_{3/2}$ is a dynamical variable in
supergravity, via its dependence on the scalar field VEVs, the
presence of such
contributions may induce, upon minimization of the effective
potential, either $m_{3/2}=0$ or $m_{3/2} \sim \mpl$. Moreover,
${\cal O} (m_{3/2}^2 \mpl^2)$ contributions to the
vacuum energy cannot be cancelled by symmetry-breaking
phenomena occurring at much lower energy scales.
This is the meaning of our third question, which forces us to go
beyond classical $N=1$ supergravity, to a more fundamental theory
where quantum gravitational effects can be consistently computed.
Today, the most natural candidate for such a theory is the heterotic
superstring [\ref{strings}], and more specifically its
four-dimensional
versions [\ref{fds}], supplemented by some assumptions on the way in
which supersymmetry is broken (existing possibilities are
non-perturbative phenomena such as gaugino condensation
[\ref{gaucond}] and tree-level string constructions such as
coordinate-dependent compactifications
[\ref{ss1}--\ref{ss3}], but there may be others). Even if
there is some arbitrariness, connected with the
assumptions on the supersymmetry-breaking mechanism, the underlying
string structure preserves some useful general properties, which
strongly restrict the form of the resulting effective supergravity
theories. For example, it is remarkable that no-scale models
naturally emerge from the string framework
[\ref{sns}, \ref{gaucond}, \ref{ss2}]. In addition,
an answer to our third question can be found in the restricted class
of string-derived no-scale supergravities
[\ref{fkpz1}--\ref{fkz},\ref{ss3}]
that are free from quadratically divergent contributions to the
one-loop vacuum energy,
proportional to ${\rm Str \,} \cm^2$. In this class of
`large-hierarchy-compatible' (LHC) models, the only corrections to
the
low-energy effective potential coming from the underlying fundamental
theory are, up to logarithms, ${\cal O} (m_{3/2}^4)$. Then, since the
only scale in the low-energy effective theory is the sliding
$m_{3/2}$, the program of inducing the hierarchy by MSSM quantum
corrections can be self-consistently carried out.

In this paper, we perform a critical reappraisal of how the hierarchy
$\msu \ll \mpl$ can be generated by MSSM quantum corrections. In
particular, we study the influence of the cosmological term of the
MSSM effective potential, which was previously neglected but plays
a very important role on the dynamical determination of
$\msu$. We find that, considering the MSSM as the low-energy limit
of a LHC supergravity model, the desired hierarchy $m_Z \sim \msu \ll
\mpl$ can be dynamically realized by the perturbative MSSM quantum
corrections.

At the level of the underlying string theory, not only the mass
parameters, but also the gauge and Yukawa couplings do indeed
depend on the VEVs of some gauge singlet fields with classically flat
potentials, called moduli, which parametrize the size and the shape
of the six-dimensional compactified space. This suggests the
following
possibility. If the dynamical mechanism that breaks supersymmetry
fixes the gauge coupling constant $\alpha_U$ to a given numerical
value at $M_U$, but leaves a residual moduli dependence of the Yukawa
couplings, then also the latter should be treated as dynamical
variables
in the low-energy effective theory. This means that the effective
potential
of the MSSM should also be minimized with respect to the moduli on
which the Yukawa couplings depend. In this paper, we discuss some
conceptual issues connected with this minimization procedure. We find
that the top-quark Yukawa coupling is dynamically driven, as a result
of the minimization with respect to the moduli fields, near its
effective infrared fixed point, corresponding to a top-quark mass
$m_t \simeq M_t^{IR} \, \sin \beta$, where $M_t^{IR} \simeq 190 \gev$
(with roughly a $10 \%$ uncertainty due to the error on $\alpha_3$,
threshold and higher-loop effects, etc.), and $(1/\sqrt{2}) < \sin
\beta < 1$ (the actual value being determined by the form of the
boundary conditions at $M_U$), thus in the range presently allowed by
experimental data [\ref{cdf}].

We conclude the paper with a quantitative example, which realizes the
above-mentioned ideas. To illustrate more clearly the important
features of the problem, we choose to work with a simplified version
of the
MSSM, leaving a more complete study to a future publication
[\ref{kprz}].

\vspace{1cm}
{\bf 2.}
The conventional treatment of radiative symmetry breaking
[\ref{radbr}]
can be briefly summarized as follows. As a starting point, one
chooses
a set of numerical input values for the independent model parameters
at
the unification scale $Q = M_U$: the soft masses
($m_0,m_{1/2},A,m_3^2$),
the superpotential mass $\mu$, the unified gauge coupling $\alpha_U$
and the third-generation\footnote{For the purposes of the present
paper, mixing effects and all other Yukawa couplings can be
neglected.}
Yukawa couplings ($\alpha_t^U,\alpha_b^U,\alpha_\tau^U$). One then
evolves all the running parameters down to a low scale $Q \sim \msu$,
according to the appropriate [\ref{rge}] renormalization group
equations (RGE), and considers the renormalization-group-improved
tree-level potential
\be
\label{vmssm}
V_0(Q) = m_1^2 v_1^2 + m_2^2 v_2^2 + 2 m_3^2 v_1 v_2 + {g'^{\,
2} + g^2 \over 8} \left( v_2^2 - v_1^2 \right)^2 + \dvc \, .
\ee
In eq.~(\ref{vmssm}), only the dependence on the real neutral
components of the Higgs fields has been kept, and $\dvc$ stands for
a Higgs-field-independent contribution to the vacuum energy
(cosmological term). All masses and coupling constants are running
parameters, evaluated at the scale $Q$. The minimization of the
potential
in eq.~(\ref{vmssm}), with respect to the dynamical variables $v_1
\equiv \langle H_1^0 \rangle$ and $v_2 \equiv \langle H_2^0 \rangle$,
is straightforward: to generate a stable minimum with non-vanishing
VEVs, one needs ${\cal B} \equiv m_1^2 m_2^2 - m_3^4 < 0$ and ${\cal
S} \equiv m_1^2 + m_2^2 - 2 |m_3^2| \ge 0$. In the determination of
the vacuum, a crucial role
is played by the large top Yukawa coupling, which strongly influences
the RGE for ${\cal B}$ and ${\cal S}$. For appropriate numerical
assignments of the boundary conditions, both ${\cal S}$ and ${\cal
B}$ are non-negative at $M_U$, but the effects of $\alpha_t$ on the
RGE
drive ${\cal B} < 0$ at scales $Q \sim \msu$, while keeping ${\cal S}
\ge 0$, and give a phenomenologically acceptable vacuum, with
$SU(2)_L \times U(1)_Y$ broken down to $U(1)_{em}$ and a mass
spectrum compatible with  present experimental data.

In this paper, we regard the MSSM as the low-energy effective theory
of an underlying supergravity model of the LHC type, where the
gravitino mass $m_{3/2}$ cannot be determined at the classical
level, due to an approximately flat direction in the space of the
scalar fields of the hidden sector, and there are no quantum
corrections to the effective potential carrying positive powers
of the cut-off scale $\mpl$. As can be shown in several examples
[\ref{fkpz1}--\ref{fkz}], this assumption can be naturally
fulfilled in a class of four-dimensional superstring models.
With the presently known mechanisms for supersymmetry breaking, in
order to allow for $m_{3/2} \ll \mpl$ one is led to
consider large values of some of the moduli fields
[\ref{larger}, \ref{ss1}--\ref{ss3}]. In this limit,
the discrete target-space duality symmetries leave as remnants some
approximate scaling symmetries for the K\"ahler metric, and one
naturally obtains approximately flat directions in moduli space,
along which the
gravitino mass can slide. Moreover, as was shown in ref.~[\ref{fkz}],
the coefficient of the ${\cal O} (m_{3/2}^2 \mpl^2)$ contributions to
the one-loop effective potential is controlled by the modular weights
associated with target-space duality, and there is a non-empty set of
LHC models in which
such a coefficient is identically vanishing.

In the present paper, we shall not consider the most general class of
LHC models, but restrict our attention to those in which some
non-perturbative dynamics fixes the VEVs of the moduli associated
with $\alpha_U$ and $M_U$,
and the boundary conditions on the explicit mass parameters of the
MSSM can
be written as\footnote{Even if $\mu$ is a globally supersymmetric
mass
parameter, several examples show that it can be related
to the scale of local supersymmetry breaking, i.e. the gravitino mass
[\ref{mu1}, \ref{ss2}, \ref{mu2}, \ref{fkz}].}
\be
\label{boundary}
m_{1/2} = \xi_1 \cdot m_{3/2} \, ,
\;\;\;\;
m_0 = \xi_2 \cdot m_{3/2} \, ,
\;\;\;\;
A = \xi_3 \cdot m_{3/2} \, ,
\;\;\;\;
m_3 = \xi_4 \cdot m_{3/2} \, ,
\;\;\;\;
\mu = \xi_5 \cdot m_{3/2} \, ,
\ee
where the scaling weights with respect to target-space duality fix
the $\xi$ parameters to constant numerical values of order one or
smaller. Then all the moduli dependence of the MSSM mass parameters
is entirely encoded in the
gravitino mass $m_{3/2}$, which should thus be considered as an extra
dynamical variable, in addition to the Higgs VEVs $v_1$ and $v_2$. If
we
take the low-energy limit and neglect the interactions of
gravitational strength, we can formally decouple the hidden sector
and recover the MSSM. Quantum effects in the underlying fundamental
theory, however, would induce a cosmological term
in the resulting MSSM effective potential;
for LHC models, this term contains no positive powers of
$\mpl$ and must therefore be proportional to $m_{3/2}^4$:
\be
\label{cosmo}
\dvc = \eta \cdot m_{3/2}^4 \, ,
\ee
obeying certain boundary conditions at $M_U$, dictated by the
structure of the hidden sector (number of states and tree-level mass
splittings):
\be
\eta(M_U) = \eta_0 \, .
\ee
We stress that, in contrast with conventional treatments, in the
present context we are forced to include the cosmological term,
since the gravitino mass is not taken as an external parameter, but
rather as a dynamical variable.

According to our program, we would like now to minimize the effective
potential of the MSSM not only with respect to the Higgs fields,
but also with respect to the new dynamical variable $m_{3/2}$,
keeping (for the moment) the values of $\alpha_U \simeq 0.04$,
$M_U \simeq 2 \times 10^{16} \gev$, $\vec{\xi}$,
$\eta_0$ and $(\atu,\alpha_b^U,\alpha_\tau^U)$
as external input data. As in the standard
approach, the role of radiative corrections will be crucial in
developing a non-zero value for the Higgs VEVs at the minimum.
Quantum corrections to the classical potential are summarized, at the
one-loop level, by the well-known formula $V_1 = V_0(Q) + \Delta V_1
(Q)$,
where\footnote{We work as usual in the 't Hooft Landau gauge and in
the mass-independent $\ov{DR}$ scheme.}
\be
\label{dv}
\Delta V_1 (Q) = {1 \over 64 \pi^2} {\rm Str} \, \cm^4 \left( \log
{\cm^2 \over Q^2} - {3 \over 2} \right) \, ,
\ee
and $V_0(Q)$ has the same functional form as the tree-level
potential, eq.~(\ref{vmssm}), but is expressed in terms of
renormalized fields
and parameters at the scale $Q$, solutions of the appropriate set of
RGE.

The RGE for the new dimensionless coupling of the theory, the
coefficient $\eta$ of the cosmological term, reads
\be
\label{etarg}
{d \eta \over d t} = {1 \over 32 \pi^2} \left[ {{\rm Str} \, \cm^4
\over m_{3/2}^4} \right]_{v_{1,2}=0} \, ,
\;\;\;\;\;
t \equiv \log Q \, .
\ee
In the MSSM,
\be
\begin{array}{c}
\left[ {\rm Str} \, \cm^4 \right]_{v_{1,2}=0}  =   4 ( m_1^4 +
m_2^4 + 2 m_3^4) + 6 ( m_{U^c_3}^4 + m_{D^c_3}^4 + 2 m_{Q_3}^4 )
+ 12 ( m_{U^c}^4 + m_{D^c}^4
\\ \\
+ 2 m_Q^4 )
+ 2 ( m_{E^c_3}^4 + 2 m_{L_3}^4 ) + 4 ( m_{E^c}^4 + 2 m_L^4 ) - 16
M_3^4 - 6 M_2^4 - 2 M_1^4 - 8 \mu^4 \, .
\end{array}
\ee
To illustrate the renormalization group evolution of the cosmological
constant term, which will play an important role in the determination
of the supersymmetry-breaking scale, we plot in fig.~1 the quantity
$(\eta-\eta_0)$, as a function of $Q$, for a number of representative
boundary conditions at $M_U$. We neglect for simplicity the effects
of $\alpha_b$ and $\alpha_{\tau}$, which can play a role only for
very large values of $\tan \beta \equiv v_2/v_1$. We can
observe that the MSSM particle content is such that $(\eta-\eta_0)$
is always driven towards negative values at sufficiently low scales,
and that the dependence on the top Yukawa coupling is not very large
within the experimentally interesting region. As we shall see in
detail
on an example, the gravitino mass dynamically relaxes to a value that
is closely related to the scale at which the coupling $\eta$
turns from positive to negative. We can then expect that the desired
hierarchy can be
generated for values of $\eta_0$ between zero and ${\cal O}(100)$,
depending
on the values of the $\vec{\xi}$ parameters, which are not in
contrast with
our present ideas on the mass splittings in the hidden sector of the
theory.

To illustrate the main point of our approach, it is convenient to
choose, as independent variables, the supersymmetry-breaking scale
$m_{3/2}^2$ and the dimensionless ratios $\hat{v}_i \equiv
(v_i/m_{3/2})$
($i=1,2$). Then the minimization condition of the one-loop effective
potential with respect to $m_{3/2}$ can be written in the form
\be
\label{master}
m_{3/2}^2 {\partial V_1 \over \partial m_{3/2}^2} = 2 V_1 + {{\rm Str
\,} \cm^4 \over 64 \pi^2} = 0 \, .
\ee
Eq.~(\ref{master}) can be interpreted as defining an infrared fixed
point for the cosmological term, since it corresponds to the
vanishing of
the associated $\beta$-function, which displays a na\"\i ve scaling
dimension
2 (with respect to $m_{3/2}^2$) and an anomalous dimension determined
by the value of ${\rm Str \,} \cm^4$. Such an equation can be
interpreted
as a non-trivial constraint to be satisfied by the MSSM parameters,
and
could hopefully discriminate among different superstring models in
which
these parameters will be eventually calculable. The minimization
conditions with respect to the variables $\hat{v}_i$ are completely
equivalent to the
ones that are usually considered in the MSSM, when the supersymmetry
breaking scale is a fixed numerical input.
A general study of the MSSM predictions, as functions of the boundary
conditions $\vec\xi$, $\eta_0$, $\alpha^U_{t,b,\tau}$, can be
performed
numerically, but goes beyond the aim of the present paper. Some
illustrative numerical results, for a particularly simple choice of
boundary conditions, will be presented in the concluding paragraph.

\vspace{1cm}
{\bf 3.}
We shall now extend the previous approach by assuming that
also $\alpha^U_{t,b,\tau}$ are dynamical variables, in analogy to
what
we did before for the supersymmetry-breaking scale $\msu$. The main
motivation for our proposal comes from superstring theory, where all
the parameters of the effective low-energy theory are related to the
VEVs of some moduli
fields, e.g. gauge-singlet fields in the hidden sector, which
after supersymmetry breaking may still
correspond to approximately flat directions at the classical level.

It is a well-known fact that, in general four-dimensional string
models, tree-level Yukawa couplings are either vanishing or of the
order of the unified gauge coupling [\ref{sns}, \ref{yukstr}].
One naturally expects the top Yukawa coupling to fall in the
latter category, so for the moment we shall concentrate on it,
assuming a tree-level relation of the form
\be
\label{hyper}
\atu = c_t \alpha_U \, ,
\ee
where $c_t$ is a model-dependent group-theoretical constant of order
unity (for example, in some fermionic constructions $c_t = 2$). At
the one-loop level, it is also well known that both gauge and Yukawa
couplings receive in general string threshold corrections
[\ref{thr}],
induced by the exchange of massive Kaluza-Klein and winding states.
Such states have masses that depend on the VEVs of some moduli
fields.
One can then consider two cases, each of which can be realized in
four-dimensional string models, as can be shown on explicit orbifold
examples.

If the top Yukawa coupling receives a string threshold correction
identical to the one of the gauge coupling, the unification condition
(\ref{hyper}) is preserved. In this case the non-perturbative
phenomena,
which we have assumed to determine $\alpha_U$, also fix the value of
$\atu$; the latter is no longer an independent parameter, and one can
perform the analysis described in the previous paragraph with one
parameter less. In particular, the structure of the RGE for
$\alpha_t$
is such that its numerical value at the electroweak scale is
always very close to the effective infrared fixed point [\ref{prh}],
$\alpha_t \simeq 1 /(4 \pi)$. Phenomenological implications of
this fact have been extensively studied in the recent literature
[\ref{fpph}].

Alternatively, eq.~(\ref{hyper}) can receive non-trivial
moduli-dependent threshold corrections, of the form
\be
\label{yuk}
{c_t \over \atu} = {1 \over \alpha_U} + F_t (T_i) \, ,
\ee
where $F_t(T_i)$ is a modular function of the singlet moduli, here
generically denoted by $T_i$. This is the situation we would like to
discuss, with both the gravitino mass $m_{3/2}$ and the top Yukawa
coupling $\atu$ depending non-trivially on the moduli fields $T_i$.
To deal with this case, we need to say something about these moduli
dependences. As for the gravitino mass, one can always choose a field
parametrization such that $m_{3/2}^2 = \alpha_U |k|^2 / V(t_V)$,
where $k$ is a superpotential term parametrizing local supersymmetry
breaking, which does not depend on the moduli corresponding to the
approximately flat directions; $V(T_i)=e^{-K}$ is the overall volume
of the internal moduli space; and $K$ is the associated K\"ahler
potential. In general, besides the overall combination of
moduli associated with the gravitino mass, to be called from now on
the `volume' modulus $t_V$, there are additional moduli. It is
plausible to assume that also some of these `shape' moduli, to be
denoted from now on by $t_i$, correspond to approximately flat
directions,
even after the inclusion of the non-perturbative physics, which
breaks
supersymmetry and fixes the value of the unified gauge coupling
constant. We shall assume here that the top Yukawa coupling $\atu$
depends on at least one of these additional moduli. Then the
minimization conditions with respect to the moduli relevant for the
low-energy theory can be written as
\be
\label{miniv}
{\partial V_1 \over \partial t_V} =
\left( {\partial V_1 \over \partial m_{3/2}^2} \right)
\left( {\partial m_{3/2}^2 \over \partial t_V} \right)
+
\left( {\partial V_1 \over \partial \alpha_t} \right)
\left( {\partial \alpha_t \over \partial \atu} \right)
\left( {\partial \atu \over \partial t_V} \right)
= 0  \, ,
\ee
\be
{\partial V_1 \over \partial t_i} =
\left( {\partial V_1 \over \partial \alpha_t} \right)
\left( {\partial \alpha_t \over \partial \atu} \right)
\left( {\partial \atu \over \partial t_i} \right)
= 0 \, .
\ee

We examine here the possible stationary points that can emerge from
the minimization with respect to the shape moduli $t_i$. There are
three generic ways of fulfilling the minimum conditions,
\be
\label{two}
{\partial \alpha_t \over \partial \alpha_t^U} = 0 \, ,
\ee
or
\be
\label{one}
{\partial \alpha_t^U \over \partial t_i} = 0 \, ,
\ee
or
\be
\label{three}
{\partial V_1 \over \partial \alpha_t} = 0 \, .
\ee

The first and most interesting possibility, eq.~(\ref{two}), always
gives a universal minimum, which corresponds to the effective
infrared fixed point. Notice that the `infrared' minima
automatically put to zero the second term in the minimization
condition with respect to the volume modulus $t_V$,
eq.~(\ref{miniv}).
One then obtains, in the large volume limit for which $\partial
m_{3/2}^2 / (\partial t_V) \ne 0$ and $m_{3/2} \ll \mpl$,
\be
m_{3/2}^2
{\partial V_1 \over \partial m_{3/2}^2} = 0
\;\;\;\;\;
{\rm and}
\;\;\;\;\;
\alpha_t \simeq \alpha_t^{IR} \, .
\ee
This universal minimum is present also if there is a non-trivial
dependence of the hidden-sector parameters on the shape moduli
$t_i$: the only additional effect would be to force those
parameters to relax to their effective infrared fixed points.

The reason of the attraction of $\alpha_t(m_{3/2})$ towards the
infrared fixed point is the particular structure of the effective
potential after the minimization with respect to the volume moduli.
Indeed, eq.~(\ref{master}) can be rewritten as
\be
V_1 |_{min} = - {1 \over 128 \pi^2}
{\rm Str \,} \cm^4  \, ,
\ee
or, more explicitly,
\be
\label{explicit}
V_1 |_{min} =  {1 \over 128 \pi^2}
(- m_t^2 C_t^2 - \ldots) \, ,
\ee
where
\be
C_t^2 = 12 \left[ m_{Q_3}^2 + m_{U_3}^2 + {m_Z^2 \over 2} \cos 2
\beta
+ \left( A_t + {\mu \over \tan \beta} \right)^2 \right] \, .
\ee
Equation~(\ref{explicit}) looks unbounded from below in the variable
$\alpha_t$. However, the actual bound is set by the effective
infrared fixed point, $\alpha_t^{IR}$. Strictly speaking, this
is excluded by the requirement of perturbative unification,
$\atu \ll 1$. However, we can say that the deepest minimum of the
effective potential corresponds to the largest value of $\alpha_t$
permitted by the structure of the moduli space of the underlying
string theory, which essentially coincides with the effective
infrared fixed point.

Equation~(\ref{explicit}) can be generalized to the case in which
also $\alpha^U_b$ and $\alpha^U_{\tau}$ are considered as dynamical
variables:
\be
\label{explicitbis}
V_1 |_{min} = {1 \over 128 \pi^2}
(- m_t^2 C_t^2 - m_b^2 C_b^2 - m_\tau^2 C_{\tau}^2 + \ldots) \, ,
\ee
where
\be
C_b^2 = 12 \left[ m_{Q_3}^2 + m_{D_3}^2 - {m_Z^2 \over 2} \cos 2
\beta
+ \left( A_b + \mu \, \tan \beta \right)^2 \right] \, ,
\ee
\be
C_{\tau}^2 = 4 \left[ m_{L_3}^2 + m_{E_3}^2 - {m_Z^2 \over 2} \cos 2
\beta
+ \left( A_\tau + \mu \, \tan \beta \right)^2 \right] \, .
\ee
Minimizing the vacuum energy tends to drive the low-energy couplings
close to an effective infrared fixed surface [\ref{kprz}]. Such a
constraint is similar (but not identical)  to the one suggested by
Nambu [\ref{nambu}], in a completely different context, as a
possibility for dynamically explaining the $m_t/m_b$ and
$m_b/m_{\tau}$
ratios. As remarked also in [\ref{bindud}], it might be possible to
adapt his approach to the present context, and work along these lines
is in progress [\ref{kprz}].

The second possibility, eq.~(\ref{one}), requires that $\partial
F(t_i) / \partial t_i = 0$, and thus involves stringy information
about the form of the $F(t_i)$ shape modular functions. Thanks to
the target-space duality symmetries [\ref{duality}], $F(t_i,t_j) =
F(t_i,1 / t_j)$, there are always such stationary points (or curves)
associated with the self-dual points of the theory ($t_i=1$). Whether
these stationary points are indeed minima depends on the details of
the string model. When they are minima, we can write the residual
moduli dependence of $\atu$ as
\be
{c_t \over \atu} = {1 \over \alpha_U} + f(t_V) \, ,
\ee
where $f(t_V) \equiv [F(T_i)]_{t_i=1}$. Then $\alpha_t$ is fixed by
the minimization with respect to $m_{3/2}$. The range of $f(t_V)$
can be estimated. If $f(t_V)$ happens to be positive and large, then
the possible minima associated with the $t_i=1$ stationary points can
be disregarded, since they give small Yukawa couplings at the scale
$\msu$, and in this case the minimum corresponding to the infrared
fixed
point is the deepest one. As discussed above, the relevant stationary
points correspond to the situation when $f(t_i)$ takes the smallest
possible value, in order to obtain the largest possible value for
$\atu/c_t$: whatever this value is, due to the infrared structure of
the RGE for $\alpha_t$, the value obtained for $\alpha_t$ at the
electroweak scale is very close to the effective infrared fixed
point.

The third possibility, eq.~(\ref{three}), depends on the low-energy
structure of the effective potential, which as a function of
$\alpha_t$ is sensitive to low-energy threshold phenomena. In
general, there may exist local stationary points of this kind, but
since the infrared fixed point is the deepest minimum that can be
obtained, we shall forget about these possible local minima in what
follows.

In summary, the possibilities described above all give $\alpha_t
\simeq \alpha_t^{IR}$ at the electroweak scale. To be more
quantitative, in
fig.~2 we plot $\alpha_t(Q)$ for some representative values of
$\atu$. We can see that, for $\atu \ge \alpha_U$, $\alpha_t$ at the
electroweak scale is always close to its effective infrared fixed
point: this gives us the numerical {\em prediction} that $m_t \sim
M_t^{IR} \sin \beta$.

\vspace{1cm}
{\bf 4.}
For simplicity, we consider here a special case of the MSSM defined
as follows:
1)~all Yukawa couplings are neglected, apart from the top-quark one,
$\alpha_t$; 2) the boundary conditions at the unification scale are
chosen
to be $10 \, m_{1/2} = m_0 = m_{3/2}$, $A = m_3^2 = \mu = 0$.
This case does not correspond to a fully realistic model, but is
suitable to
illustrate in a simple way the main conceptual points of our
approach. We would like to stress, however, that our considerations
can be easily generalized to the
fully-fledged MSSM [\ref{kprz}]. Since, in the special case under
consideration, $m_3^2=0$ and $m_1^2 > 0$ at all scales $Q < M_U$, as
can be verified by looking at the general structure of the RGE
[\ref{rge}],
we can assume $v_1=0$ at the minimum, and restrict our attention to
the
dependence of the RG-improved tree-level potential on $v \equiv v_2$,
\be
\label{vtree}
V_0 = m_2^2 v^2 + \frac{g^2+g'\,^2}{8} v^4 + \eta \,  m_{3/2}^4 = {1
\over 8
\pi \alpha_Z} \left( 4 m_2^2 m_Z^2 + m_Z^4 \right) + \eta  \,
m_{3/2}^4 \, ,
\ee
where to write the second expression we have used the definition
$\alpha_Z \equiv (g^2+g'\,^2) / (4 \pi)$ and the tree-level relation
$m_Z^2 = (g^2 + g' \,^2) v^2 / 2$. In the following, we shall always
choose $Q = 2 \, m_{3/2}$, which will give an approximation good
enough
for the present purposes [\ref{grz}]. It is however clear that, if
one
wants to extend the present work to the fully-fledged MSSM and give
accurate
quantitative predictions, threshold effects have to be more
accurately
modelled [\ref{kprz}].

The potential of eq.~(\ref{vtree}) is easily studied. The stability
condition along the $v_1=v_2$ direction is just $m_1^2 + m_2^2 \ge
0$, and minimization with respect to $m_Z$ (or, equivalently, with
respect to $v$) gives
\be
\label{mzimin}
\begin{array}{ccl}
m_Z^2 = - 2 m_2^2
&
{\rm if}
&
m_2^2 < 0 \, ,
\\
m_Z^2 = 0
&
{\rm if}
&
m_2^2 > 0 \, .
\end{array}
\ee
Substituting this into eq.~(\ref{vtree}), we get
\be
\label{vafter}
\begin{array}{ccl}
\displaystyle{
V_0 = \left( \eta - {1 \over 2 \pi \alpha_Z} {m_2^4 \over m_{3/2}^4}
\right) m_{3/2}^4}
&
{\rm if}
&
m_2^2 < 0 \, ,
\\
V_0 = \eta \, m_{3/2}^4
&
{\rm if}
&
m_2^2 > 0 \, .
\end{array}
\ee
The two RGEs that control the dynamical determination of $m_Z$ and
$m_{3/2}$ are the one for the cosmological parameter $\eta$,
equation~(\ref{etarg}), and the one for the mass parameter $m_2^2$,
which reads
in our simplified case:
\be
\label{mtwo}
{d m_2^2 \over d t} = {1 \over 2 \pi} \left[ - 3 \alpha_2 M_2^2 -
\alpha' M_1^2 + 3 \alpha_t \left( m_{Q_3}^2 +m_{U^c_3}^2 + m_2^2 +
A_t^2 \right) \right] \, .
\ee
The results of the minimization of the effective potential are
illustrated in fig.~3, which shows contours of constant $m_Z$,
$m_{3/2}$ and $m_t/m_Z$ in the $(\eta_0,\atu)$ plane. For
$\eta_0 \sim 9$, the transmutation scale at which $\eta$
turns from positive to negative is close to the electroweak
scale, and the experimentally observed value of $m_Z$ can be
reproduced. At these minima, the gravitino mass is also of order
$m_Z$, and $m_t \simeq (1.8$--$2) \, m_Z$ can be easily reproduced
for an appropriate choice of $\atu$.

\vspace{1cm}
{\bf 5.}
To conclude, we summarize the general features of the dynamical
determination of ($m_Z,m_{3/2},m_t$), when one regards the MSSM
as the low-energy limit of one of the LHC supergravity models
considered in ref.~[\ref{fkz}].

The LHC models were defined as the $N=1$ supergravities where, in the
presence
of well-defined quantum gravitational corrections, the contributions
to the vacuum energy are not larger than ${\cal O}(m_{3/2}^4)$. At
least at the one-loop level, examples of LHC models can be obtained
from four-dimensional superstring models where supersymmetry is
spontaneously broken at the tree level, e.g. by coordinate-dependent
compactifications. Other candidate LHC models are certain superstring
effective supergravities, where supersymmetry breaking is induced by
non-perturbative phenomena such as gaugino condensation.
In LHC models, the effective potential is cut-off-independent, up to
benign logarithmic corrections, and the scale is set by the gravitino
mass $m_{3/2}$, which in turn depends on a singlet modulus field
$t_V$, corresponding to an approximately flat direction. In
particular, we concentrated here on the case in which the only
moduli dependence of the MSSM mass terms is via
$m_{3/2}$. In LHC models, minimization of the effective potential $V$
with respect to the volume modulus $t_V$ sets $m_{3/2}$ to a value
defined
by the infrared fixed point of the vacuum energy $m_{3/2}^2
( \partial V / \partial m_{3/2}^2) = 0 $. We found that, for a wide
and reasonable range of coupling constants
$(\alpha_U,\alpha^U_{t,b,\tau})$ and other dimensionless parameters
$(\vec{\xi},\eta_0)$, this corresponds to the
desired hierarchy $m_Z \sim m_{3/2} \ll \mpl$.

Inspired by four-dimensional superstring models, we examined the case
in which also the Yukawa couplings are field-dependent dynamical
variables, concentrating on the largest one, associated with the top
quark. In analogy with the previous result on $m_{3/2}$, we found
that also the top Yukawa coupling $\alpha_t$ is driven to its
effective infrared fixed point. This left us
with a definite prediction for the top-quark mass, $m_t=M_t^{IR} \,
\sin \beta$, where $M_t^{IR} \sim 190 \gev$ and $\sin \beta$ can be
computed in terms of the $(\vec{\xi},\eta_0)$ parameters, derivable
from the fundamental theory at the cut-off scale. A more general
analysis of mass hierarchies, including not only $m_{3/2}/M_U$,
$m_Z/M_U$ and $m_t/M_U$,
but also some fermion mass ratios such as $m_b/m_t$ and
$m_{\tau}/m_t$, appears to be feasible in the framework of LHC models
and is currently under study
[\ref{kprz}].

\section*{Acknowledgements}
One of us (F.Z.) would like to thank the ITP at Santa Barbara
for the kind hospitality during the final stage of this work.

\newpage
\section*{References}
\begin{enumerate}
\item
\label{mssm}
P.~Fayet, Phys. Lett. B69 (1977) 489 and B84 (1979) 416;
\\
S.~Dimopoulos and H.~Georgi, Nucl. Phys. B193 (1981) 150.
\\
For reviews and further references, see, e.g.:
\\
H.-P. Nilles, Phys. Rep. 110 (1984) 1;
\\
S. Ferrara, ed., `Supersymmetry' (North-Holland, Amsterdam, 1987);
\\
F.~Zwirner, in `Proceedings of the 1991 Summer School in High Energy
Physics and Cosmology', Trieste, 17 June--9 August 1991 (E.~Gava,
K.~Narain, S.~Randjbar-Daemi, E.~Sezgin and Q.~Shafi, eds.), Vol.~1,
p.~193.
\item
\label{radbr}
L.E.~Ib\'a\~nez and G.G.~Ross, Phys. Lett. B110 (1982) 215;
\\
L.~Alvarez-Gaum\'e, M.~Claudson and M.B.~Wise, Nucl. Phys. B207
(1982) 96;
\\
K.~Inoue, A.~Kakuto, H.~Komatsu and S.~Takeshita, Prog. Theor. Phys.
68 (1982) 927 and 71 (1984) 413;
\\
J.~Ellis, D.V.~Nanopoulos and K.~Tamvakis, Phys. Lett. B121 (1983)
123;
\\
L.~Alvarez-Gaum\'e, J.~Polchinski and M.B.~Wise, Nucl. Phys. B221
(1983) 495;
\\
L.E.~Ib\'a\~nez and C.~Lopez, Nucl. Phys. B233 (1984) 511;
\\
C.~Kounnas, A.B.~Lahanas, D.V.~Nanopoulos and M.~Quir\'os, Nucl.
Phys. B236 (1984) 438.
\item
\label{cfgvp}
E.~Cremmer, S.~Ferrara, L.~Girardello and A.~Van~Proeyen, Phys. Lett.
B116 (1982) 231 and Nucl. Phys. B212 (1983) 413;
\\
J.~Bagger and E.~Witten, Phys. Lett. B118 (1982) 103;
\\
J.~Bagger, Nucl. Phys. B211 (1983) 302.
\item
\label{bfs}
A.H.~Chamseddine, R.~Arnowitt and P.~Nath, Phys. Lett. 49 (1982) 970;
\\
R.~Barbieri, S.~Ferrara and C.A.~Savoy, Phys. Lett. B119 (1982) 343;
\\
E.~Cremmer, P.~Fayet and L.~Girardello, Phys. Lett. B122 (1983) 41;
\\
L.~Hall, J.~Lykken and S.~Weinberg, Phys. Rev. D27 (1983) 2359;
\\
S.K.~Soni and H.A.~Weldon, Phys. Lett. B126 (1983) 215.
\item
\label{cfkn}
E.~Cremmer, S.~Ferrara, C.~Kounnas and D.V.~Nanopoulos, Phys. Lett.
B133 (1983) 61.
\item
\label{ekn}
J.~Ellis, A.B.~Lahanas, D.V.~Nanopoulos and K.~Tamvakis, Phys. Lett.
B134 (1984) 429;
\\
J.~Ellis, C.~Kounnas and D.V.~Nanopoulos, Nucl. Phys. B241 (1984) 406
and B247 (1984) 373.
\item
\label{strings}
D.J.~Gross, J.A.~Harvey, E.~Martinec and R.~Rohm, Phys. Rev. Lett. 54
(1985)
502; Nucl. Phys. B256 (1985) 253 and B267 (1985) 75.
\\
For a review and further references see, e.g.:
\\
M.B.~Green, J.H.~Schwarz and E.~Witten, {`Superstring theory'}
(Cambridge University Press, Cambridge, 1987).
\item
\label{fds}
P.~Candelas, G.~Horowitz, A.~Strominger and E.~Witten, Nucl. Phys.
B258 (1985) 46;
\\
L.~Dixon, J.~Harvey, C.~Vafa and E.~Witten, Nucl. Phys. B261 (1985)
678;
\\
K.S.~Narain, Phys. Lett. B169 (1986) 41;
\\
K.S.~Narain, M.H.~Sarmadi and E.~Witten, Nucl. Phys. B279 (1987) 369;
\\
W.~Lerche, D.~L\"ust and A.N.~Schellekens, Nucl. Phys. B287 (1987)
477;
\\
H.~Kawai, D.C.~Lewellen and S.-H.H. Tye, Nucl. Phys. B288 (1987) 1;
\\
I.~Antoniadis, C.~Bachas and C.~Kounnas, Nucl. Phys. B289 (1987) 87.
\\
K.S.~Narain, M.H.~Sarmadi and C.~Vafa, Nucl. Phys. B288 (1987) 551.
\\
For a review and further references see, e.g.:
\\
B.~Schellekens, ed., {`Superstring construction'} (North-Holland,
Amsterdam, 1989).
\item
\label{gaucond}
J.-P.~Derendinger, L.E.~Ib\'a\~nez and H.P.~Nilles, Phys. Lett. B155
(1985) 65;
\\
M.~Dine, R.~Rohm, N.~Seiberg and E.~Witten, Phys. Lett. B156 (1985)
55;
\\
C.~Kounnas and M.~Porrati, Phys. Lett. B191 (1987) 91.
\item
\label{ss1}
R.~Rohm, Nucl. Phys. B237 (1984) 553;
\\
C.~Kounnas and M.~Porrati, Nucl. Phys. B310 (1988) 355;
\\
C.~Kounnas and B.~Rostand, Nucl. Phys. B341 (1990) 641.
\item
\label{ss2}
S.~Ferrara, C.~Kounnas, M.~Porrati and F.~Zwirner,
Nucl. Phys. B318 (1989) 75;
\\
M.~Porrati and F.~Zwirner, Nucl. Phys. B326 (1989) 162;
\\
I.~Antoniadis and C.~Kounnas, Phys. Lett. B261 (1991) 369.
\item
\label{ss3}
I.~Antoniadis, Phys. Lett. B246 (1990) 377.
\item
\label{sns}
E.~Witten, Phys. Lett. B155 (1985) 151;
\\
S.~Ferrara, C.~Kounnas and M.~Porrati, Phys. Lett. B181 (1986) 263;
\\
S.~Ferrara, L.~Girardello, C.~Kounnas and M.~Porrati, Phys. Lett.
B193 (1987) 368;
\\
I.~Antoniadis, J.~Ellis, E.~Floratos, D.V.~Nanopoulos and T.~Tomaras,
Phys. Lett. B191 (1987) 96;
\\
S.~Ferrara, L.~Girardello, C.~Kounnas and M.~Porrati, Phys. Lett.
B194 (1987) 358;
\\
M.~Cvetic, J.~Louis and B.~Ovrut, Phys. Lett. B206 (1988) 227;
\\
S.~Ferrara and M.~Porrati, Phys. Lett. B216 (1989) 1140;
\\
M.~Cvetic, J.~Molera and B.~Ovrut, Phys. Rev. D40 (1989) 1140;
\\
L.~Dixon, V.~Kaplunovsky and J.~Louis, Nucl. Phys. B329 (1990) 27.
\item
\label{fkpz1}
S.~Ferrara, C.~Kounnas, M.~Porrati and F.~Zwirner, Phys. Lett. B194
(1987) 366.
\item
\label{add}
C.~Kounnas, in `Properties of SUSY Particles', Proceedings of the
23rd Workshop on the INFN Eloisatron Project, Erice,
28~September--4~October 1992 (L.~Cifarelli and V.A.~Khoze eds.),
World Scientific, Singapore, 1992 p.~496;
\\
I.~Antoniadis, C.~Mu\~noz and M.~Quir\'os, Nucl. Phys. B397 (1993)
515.
\item
\label{fkz}
S.~Ferrara, C.~Kounnas and F.~Zwirner, preprint CERN-TH.7192/94,
LPTENS-94/12, UCLA/94/TEP13, hep-th/9405188.
\item
\label{cdf}
G.~Altarelli, preprint CERN-TH.7045/93, Plenary talk given at the
International Europhysics
Conference on High Energy Physics, Marseille, 22--28 July 1993,
 to appear in the Proceedings, and
references therein;
\\
F.~Abe et al. (CDF Collaboration), preprint FERMILAB-PUB-94/097-E.
\item
\label{kprz}
C.~Kounnas, I.~Pavel, G.~Ridolfi and F.~Zwirner, work in progress.
\item
\label{rge}
R. Barbieri, S. Ferrara, L. Maiani, F. Palumbo and C.A. Savoy, Phys.
Lett. B115 (1982) 212;
\\
K. Inoue, A. Kakuto, H. Komatsu and S. Takeshita, as in
ref.~[\ref{radbr}];
\\
L.~Alvarez-Gaum\'e, J.~Polchinski and M.B.~Wise, as in
ref.~[\ref{radbr}];
\\
J.-P.~Derendinger and C.A.~Savoy, Nucl. Phys. B237 (1984) 307.
\item
\label{larger}
R.~Rohm and E.~Witten, Ann. Phys. 170 (1986) 454;
\\
M.~Dine and N.~Seiberg, Nucl. Phys. B301 (1988) 357;
\\
T.~Banks and L.J.~Dixon, Nucl. Phys. B307 (1988) 93;
\\
I.~Antoniadis, C.~Bachas, D.~Lewellen and T.N.~Tomaras, Phys. Lett.
B207 (1988) 441.
\item
\label{mu1}
S.K.~Soni and H.A.~Weldon, as in ref.~[\ref{bfs}];
\\
G.F.~Giudice and A.~Masiero, Phys. Lett. B206 (1988) 480.
\item
\label{mu2}
I.~Antoniadis, C.~Mu\~noz and M.~Quir\'os, as in ref.~[\ref{add}];
\\
V.~Kaplunovsky and J.~Louis, Phys. Lett. B306 (1993) 269;
\\
A.~Brignole, L.E.~Ib\'a\~nez and C.~Mu\~noz, Madrid preprint
FTUAM-26/93;
\\
I. Antoniadis, E. Gava, K.S. Narain and T.R. Taylor, Northeastern
University  \\ preprint NUB-3084 (1994).
\item
\label{yukstr}
A.~Strominger, Phys. Rev. Lett. 55 (1985) 2547;
\\
A.~Strominger and E.~Witten, Commun. Math. Phys. 101 (1985) 341;
\\
L.J.~Dixon, E.~Martinec, D.~Friedan and S.~Shenker, Nucl. Phys.
B282 (1987) 13;
\\
P.~Candelas, Nucl. Phys. B298 (1988) 458;
\\
B.R.~Greene, C.A.~L\"utken and G.G.~Ross, Nucl. Phys. B325 (1989)
101;
\\
B.R.~Greene, K.H.~Kirklin, P.J.~Miron and G.G.~Ross, Phys. Lett.
B192 (1989) 111.
\item
\label{thr}
V.S.~Kaplunovsky, Nucl. Phys. B307 (1988) 145;
\\
L.J.~Dixon, V.S.~Kaplunovsky and J.~Louis, Nucl. Phys. B355 (1991)
649;
\\
J.-P.~Derendinger, S.~Ferrara, C.~Kounnas and F.~Zwirner, Nucl. Phys.
B372 (1992) 145;
\\
G.~Lopez Cardoso and B.A.~Ovrut, Nucl. Phys. B369 (1992) 351;
\\
I.~Antoniadis, K.S.~Narain and T.~Taylor, Phys. Lett. B276 (1991) 37;
\\
I.~Antoniadis, E.~Gava, K.S.~Narain and T.R.~Taylor, Nucl. Phys. B407
(1993) 706.
\item
\label{prh}
B. Pendleton and G.G. Ross, Phys. Lett. B98 (1981) 291;
\\
C. Hill, Phys. Rev. D24 (1981) 691;
\\
L.~Alvarez-Gaum\'e, J.~Polchinski and M.B.~Wise, as in
ref.~[\ref{radbr}];
\\
J. Bagger, S. Dimopoulos and E. Mass\'o, Phys. Rev. Lett. 55 (1985)
920.
\item
\label{fpph}
M.~Carena, T.E.~Clark, C.E.M.~Wagner, W.A.~Bardeen and K.~Sasaki,
Nucl. Phys. B369 (1992) 33;
\\
H.~Arason, D.J.~Casta\~no, B.~Keszthelyi, S.~Mikaelian, E.J.~Piard,
P.~Ramond and B.D.~Wright, Phys. Rev. Lett. 67 (1991) 2933;
\\
S.~Kelley, J.L.~Lopez and D.V.~Nanopoulos, Phys. Lett. B278 (1992)
140;
\\
V.~Barger, M.S.~Berger, P.~Ohmann and R.J.N.~Phillips, Phys. Lett.
B314 (1993) 351;
\\
M.~Carena, S.~Pokorski and C.E.M.~Wagner, Nucl. Phys. B406 (1993) 59;
\\
W.A.~Bardeen, M.~Carena, S.~Pokorski and C.E.M.~Wagner, Phys. Lett.
B320 (1994) 110;
\\
M.~Carena, M.~Olechowski, S.~Pokorski and C.E.M.~Wagner, Nucl. Phys.
B419 (1994) 213;
\\
P.~Langacker and N.~Polonsky, Phys. Rev. D49 (1994) 1454.
\item
\label{nambu}
Y.~Nambu, as quoted by T.~Gherghetta at the IFT Workshop on Yukawa
Couplings, Gainesville, February 11--13, 1994, to appear in the
Proceedings.
\item
\label{bindud}
P.~Bin\'etruy and E.~Dudas, Orsay preprint LPTHE 94/35.
\item
\label{duality}
K.~Kikkawa and M.~Yamasaki, Phys. Lett. B149 (1984) 357;
\\
N.~Sakai and I.~Senda, Progr. Theor. Phys. 75 (1986) 692;
\\
V.P.~Nair, A.~Shapere, A.~Strominger and F.~Wilczek, Nucl. Phys.
B287 (1987) 402;
\\
B.~Sathiapalan, Phys. Rev. Lett. 58 (1987) 1597;
\\
R.~Dijkgraaf, E.~Verlinde and H.~Verlinde, Commun. Math. Phys. 115
(1988) 649;
\\
A.~Giveon, E.~Rabinovici and G.~Veneziano, Nucl. Phys. B322 (1989)
167;
\\
A.~Shapere and F.~Wilczek, Nucl. Phys. B320 (1989) 301;
\\
M.~Dine, P.~Huet and N.~Seiberg, Nucl. Phys. B322 (1989) 301.
\\
For a review and further references see, e.g.:
\\
A.~Giveon, M.~Porrati and E.~Rabinovici, preprint RI-1-94,
NYU-TH-94-01-01, to appear in Physics Reports.
\item
\label{grz}
G.~Gamberini, G.~Ridolfi and F.~Zwirner, Nucl. Phys. B331 (1990) 331.
\end{enumerate}
\vfill{
\section*{Figure captions}
\begin{itemize}
\item[Fig.1:]
Renormalization of the cosmological term of the MSSM for some
representative boundary conditions at $M_U$: a) $m_{1/2}=m_{3/2}$,
$m_0=A=m_3=\mu=0$; \\ b) $m_0=m_{3/2}$, $m_{1/2}=A=m_3=\mu=0$; c)
$m_{1/2}=m_0=m_{3/2}$, $A=m_3=\mu=0$;
d) $m_{1/2}=m_0=A=m_3=\mu=m_{3/2}$. For simplicity, the top Yukawa
coupling has been assigned the representative values $\atu=0$
(dashed lines), $\atu=\alpha_U$
(solid lines), $\atu = 4 \alpha_U$ (dash-dotted lines), and all other
Yukawa
couplings have been neglected.
\item[Fig.2:]
The running top Yukawa coupling $\alpha_t(Q)$, in the interval $m_Z
\le Q \le M_U$, computed at the one-loop level and neglecting
threshold effects, for some representative choices of the boundary
condition at the unification scale:
$\atu = \alpha_U$ (solid line); $\atu = 2 \alpha_U$ (dashed
line);
$\atu = 4 \, \alpha_U$ (dash-dotted line).
\item[Fig.3:]
Contours of constant $m_Z = 60, 90, 120 \gev$ (solid lines),
$m_0=10 \, m_{1/2} = m_{3/2} = 100, 300, 500 \gev$  (dashed lines)
and $m_t/m_Z=1.8,1.9,2$ (dash-dotted lines), in the $(\eta_0,\atu)$
plane, for the simplified version of the MSSM discussed in the text.
\end{itemize}
}
\end{document}